\documentclass[prd,aps]{revtex4}
\usepackage{graphicx}

\begin{document}

\title{Well posed constraint-preserving boundary conditions\\ for 
the linearized Einstein equations}

\author{Gioel Calabrese, Jorge Pullin, Olivier Sarbach, Manuel Tiglio}

\affiliation{Department of Physics and Astronomy, Louisiana State
University, 202 Nicholson Hall, Baton Rouge, Louisiana 70803-4001}

\author{Oscar Reula} 
\affiliation{Facultad de Matem\'atica, Astronom\'{\i}a y F\'{\i}sica,
Universidad Nacional de C\'ordoba,\\ Ciudad
Universitaria, 5000 C\'ordoba, Argentina.}

\begin{abstract}

In the Cauchy problem of general relativity one considers initial data
that satisfies certain constraints. The evolution equations guarantee
that the evolved variables will satisfy the constraints at later 
instants of time. This is only true within the domain of dependence of
the initial data. If one wishes to consider situations where the evolutions
are studied for longer intervals than the size of the domain of dependence,
as is usually the case in three dimensional numerical relativity, one
needs to give boundary data. The boundary data should be specified in 
such a way that the constraints are satisfied everywhere, at all times.
In this paper we address this problem for the case of general relativity
linearized around Minkowski space using the generalized Einstein-Christoffel 
symmetric hyperbolic system of evolution equations.
We study the evolution equations for the constraints, specify boundary 
conditions for them that make them well posed  and further choose 
these boundary conditions in such a way that the evolution equations 
for the metric variables are also well posed. We also consider
the case of a manifold with a non-smooth boundary, as is the usual case
of the cubic boxes commonly used in numerical relativity. The
techniques discussed
should be applicable to more general cases, as linearizations
around more complicated backgrounds, and may be used to establish
well posedness  in the full non-linear case. 
\end{abstract}

\maketitle
\section{Introduction}

In the Cauchy problem of general relativity, Einstein's equations
split into a set of evolution equations and a set of constraint
equations. Given initial data that satisfies the constraints one can
show that the evolution equations imply that the constraints continue
to hold in the domain of dependence of the initial slice.  However, in
numerical relativity, due to finite computer resources, one usually
solves the field equations in a domain of the form $(t,x^i) \in {\cal
R}\times\Omega$, with $\Omega$ a bounded space-like surface with
boundary $\partial \Omega$.  In this case the domain of dependence of
the initial slice is ``too small'' and one wants to evolve beyond it.
Then the question that arises is what boundary conditions to give at
the artificial boundary ${\cal R}\times \partial\Omega$ .  This
problem has two aspects to it. First, one wishes to prescribe boundary
conditions that make the problem a well posed one which preserves the
constraints. In addition to that, one might desire to embed in the
boundary conditions some physically appealing property (for instance
that no gravitational radiation enters the domain). These two aspects
are in principle separate. In this paper we will concentrate on the
first one, namely, how to prescribe consistent boundary
conditions. The construction we propose ends up giving ``Dirichlet or
Neumann-like'' boundary conditions on some components of the metric,
with some free sources.  The latter allow freedom to consider boundary
conditions for any spacetime in any slicing.

If the field equations are cast into a first order symmetric
hyperbolic form there is a known prescription for achieving well
posedness for an initial-boundary value problem (this is discussed in
detail in section II), namely maximal dissipative boundary
conditions \cite{laxfr}. Well posedness is a necessary condition for
implementing a stable numerical code in the sense of Lax's theorem
\cite{convergence}.  However, additional work is needed if one wishes
to have a code that is not only stable but that preserves the
constraints throughout the domain of evolution.  As we stated above,
this implies giving initial and boundary data that guarantees that the
constraints are satisfied. Traditionally, most numerical relativity
treatments have been careful to impose initial data that satisfies the
constraints. However, very rarely boundary conditions that lead to
well posedness are used, and much less frequently are they consistent
with the constraints. Only recently, following Friedrich and Nagy
\cite{fn}, has work with numerical relativity in mind started along
these lines
\cite{stewart},\cite{iriondo},\cite{szilagyi1},\cite{bardeen},
\cite{calabrese}. Of particular interest is the recent paper of
Szilagyi and Winicour \cite{szilagyi2}, which outlines a construction
with several points in common with the one we describe here.

In this article we derive well posed, constraint-preserving boundary
 conditions for the generalized \cite{kst} Einstein-Christoffel (EC)
 \cite{ay} 
 symmetric hyperbolic formulations of Einstein's equations, when linearized
 around flat spacetime.  The procedure consists in studying the
 evolution system for the constraints and making sure that the
 corresponding initial-boundary value problem is well-posed through
 appropriate boundary conditions. There is some freedom in the
 specification of the latter.  Next, we translate them into boundary
 conditions for the variables of the main evolution system. The main 
 difficulty consists in making use of the freedom that one has in the
 boundary conditions for the evolution system of the constraints in
 such a way that the resulting boundary conditions for the main
 evolution system ensure well-posedness. We consider the case of a
 non-smooth cubic boundary as is of interest in numerical relativity.
 This involves the additional complication of being careful about
 ensuring compatibility at the edges joining the faces. To our
 knowledge, this is the first detailed analysis of the
 initial-boundary value problem with a non-smooth boundary in the
 gravitational context.

The organization of this paper is as follows: in section \ref{energy}
we review the basic technique of energy estimates used to prove well
posedness. In section \ref{kst} we review the generalized EC symmetric
hyperbolic formulation, write down the evolution
system for the constraints and analyze under what conditions the
latter is symmetric hyperbolic. 
In section \ref{cpbc} we compute the characteristic variables for the
main evolution system and for the system that evolves the constraints.
These characteristic variables are a key element for prescribing the
boundary conditions that lead to well-posedness. In section
\ref{cpbcs} we write down the constraint preserving boundary conditions
in terms of the variables of the main evolution system. In section
\ref{well_posedness} we derive the necessary energy estimates to show that  all
the systems involved in our constraint-preserving treatment are 
well posed. In section \ref{summary} we summarize the
constraint-preserving construction of this paper. 
We end with a discussion of possible
future improvements to and applications of the boundary 
conditions introduced in this paper.

\section{Basic energy estimates}
\label{energy}

In this section we review some basic notions of energy estimates for
systems of partial differential equations, as discussed, for instance, 
in \cite{kreiss1}. Consider a first order in time and space linear evolution
system of the form
\begin{equation}
\partial_t u = A^j\partial_j u,
\label{Eq:MainEqs}
\end{equation}
where $u = u(t,x^i)$ is a vector valued function, $t\geq 0$, $x^i \in
\Omega$, and where the matrices $A^1$, $A^2$ and $A^3$ are
constant. The symbol $P(\vec{n})$ is defined as $P:= A^i n_i$. The
system is symmetric (or symmetrizable) hyperbolic if there is a
symmetrizer for $P$. That is, a positive definite, Hermitian matrix
$H$ {\em independent of $n^i$} such that $H P = P^{\dagger }H$ for all
$n_i$ with $\sum _{i=1\ldots 3} n_i^2=1$.

In order to show well-posedness of the initial value problem for such a system, 
one usually derives a bound for the energy
\begin{equation}
E(t) = \int_\Omega (u,Hu)\, d^3 x\; ,
\label{Eq:Energy}
\end{equation}
where $(.,.)$ denotes the standard scalar product.
Taking a time derivative of (\ref{Eq:Energy}), using equation (\ref{Eq:MainEqs}),
the fact that $H A^j$ are symmetric matrices, and Gauss' theorem, one obtains
$$
\frac{d}{dt}\, E(t) = \int_\Omega 2(u, H A^j\partial_j u)\, d^3 x
 = \int_\Omega \partial_j (u, H A^j u)\, d^3 x
 = \int_{\partial\Omega} (u, H P(\vec{n}) u)\, d\sigma,
$$
where $\vec{n}$ is now the unit outward normal to the boundary $\partial\Omega$
of the domain. For simplicity let's assume that $P(\vec{n})$ has only the eigenvalues 
$\pm 1$ and $0$.
Let $u^{(+)}$, $u^{(-)}$, and $u^{(0)}$ denote the projections of $u$ 
onto the eigenspaces corresponding to the eigenvalues $1$, $-1$
and $0$, respectively. That is, $u=u^{(+)}+u^{(-)}+u^{(0)}$ and 
$P(\vec{n})u = u^{(+)} - u^{(-)}$. 
Then  
$$
(u, HP(\vec{n})u) = (u^{(+)},Hu^{(+)}) - (u^{(-)}, Hu^{(-)}).
$$
If we impose boundary conditions of the form $u^{(+)} = R u^{(-)}$
with $R$ ``small enough'' so that $R^TH R \leq H$ \footnote{In the
sense that 
precisely $(u,R^TH R u) \leq (u,H u)$ $\forall u$.}, it follows that
$E(t) \leq E(0)$ for all $t\geq 0$. These kind of energy estimates are
a key ingredient in well-posedness proofs.  One can generalize
this result to boundary conditions of the form
\begin{equation}
u^{(+)} = R u^{(-)} + g,
\label{Eq:MaxDiss}
\end{equation}
where $g = g(t,x^A)$ is a prescribed function at the boundary
and $R^T H R \leq H$, as before. In this case, an energy estimate can
be obtained as follows: We first choose a function $\psi$ that
satisfies $\psi^{(+)} = R \psi^{(-)} + g$ at the boundary.
Then, we consider the variable $\tilde{u} \equiv u - \psi$
instead of $u$ which now satisfies the homogeneous boundary
condition $\tilde{u}^{(+)} = R \tilde{u}^{(-)}$ and the
modified evolution equation
$$
\partial_t\tilde{u} = A^j\partial_j \tilde{u} + F,
$$
where $F = A^j\partial_j\psi - \partial_t\psi$ is a forcing term.
One now repeats the estimate for the energy defined 
in (\ref{Eq:Energy}), with $u$ replaced by $\tilde{u}$, 
and obtains
$$
\frac{d}{dt}\, E(t) \leq 2\int_\Omega (\tilde{u},H F) d^3 x 
 \leq 2\| \tilde{u} \| \cdot \| F \|,
$$
where $\| \tilde{u} \| \equiv E^{1/2}$, 
$\| F \| \equiv (\int (F,H F)d^3 x)^{1/2}$ and
where we have used Schwarz's inequality.
Therefore, we obtain the estimate
$$
\| \tilde{u}(t,.) \| \leq \| \tilde{u}(0,.) \| + \int_0^t \| F(s,.) \| ds.
$$

Similar estimates can be obtained for systems with non-constant
coefficients, as is the case, for instance, of
linearizations around a given non-Minkowski metric, as in black hole
perturbations. In the non-linear case the
proofs of bounds are only for finite amounts of time, since 
solutions can blow up in a finite time starting 
from initial data with finite energy.

The existence of an energy estimate implies that the initial-boundary
value problem is well posed. By this we mean that there exists a unique
smooth solution to the problem for which the energy estimate holds
\cite{secchi,rauch}. It should be noted that these results are in principle 
only valid for smooth boundaries. Later in this paper we will discuss
boundaries that are non-smooth. It should be understood that in those
cases the energy estimates we derive do not necessarily imply the
existence and uniqueness of smooth solutions.

\section{The field equations and their constraint propagation}
\label{kst}

In this section we present the field equations that we will use through
this paper, and analyze the constraints propagation.
In the first subsection we present the evolution equations for the
main variables within the generalized Einstein-Christoffel symmetric
hyperbolic formulation of Einstein's equations. 
In the second subsection we analyze, in the fully non-linear case, the
evolution equations for the constraints within this  
formulation and derive necessary conditions for the latter to be
symmetrizable. We will need this system to be symmetric hyperbolic in
order to derive an energy estimate in the way we sketched in Section
\ref{energy}. Rather surprisingly, it turns out that in the original
EC system the constraints' propagation does not seem to be
symmetrizable. Imposing the symmetric hyperbolicity condition naturally restricts the free
parameter of the system to an open interval. As a side note, there
seems to be some correlation between the stability properties of the
system found in numerical experiments and this natural choice.

\subsection{The formulation}

In \cite{kst} the following symmetric hyperbolic
system \footnote{This system corresponds to System 3 of \cite{kst} with
their parameter $\hat{z}$ set to zero.}
of evolution equations 
for the three-metric ($g_{ij}$), the extrinsic curvature ($K_{ij}$), and
some extra variables $f_{kij}$ that are introduced in order to make
the system first order in space is derived:
\begin{eqnarray}
\partial_0 g_{ij} &=& -2 K_{ij} \, , 
\label{Eq:gij}\\
\partial _0 K_{ij} &=& -\partial^k f_{kij}  + l.o. \, , 
\label{Eq:Kij}\\
\partial _0 f_{kij} &=& -\partial _k K_{ij} + l.o. \, , 
\label{Eq:fkij}
\end{eqnarray}
where $\partial^k \equiv g^{kl}\partial_l$ and where 
$\partial_0 = (\partial_t - \pounds_{\beta})/N$ is the
derivative operator along the normal to the spatial $t=const.$ slices.
The shift $\beta^i$ is an a priori prescribed function on spacetime  
while the lapse $N$ is determined by $N = g^{1/2}\, e^Q$, with $Q$ 
 a priori prescribed. Here and in the following
$l.o.$ stands for ``lower order terms''. These terms depend on $g_{ij}$,
$K_{ij}$, $f_{kij}$, $Q$ and $\beta^i$ but not on the derivatives of
$g_{ij}$, $K_{ij}$ or $f_{kij}$.
Provided that the constraints (see equation (\ref{Eq:Constraints}) below)
are satisfied, the spatial derivatives, $d_{kij} \equiv \partial_k g_{ij}$, 
of the three-metric are obtained from
\begin{equation}
d_{kij} = 2\, f_{kij} + \eta\, g_{k(i} \left( f_{j)s}^{\;\;\;\;\; s} -
        f^s_{\; j)s} \right) + \frac{\eta-4}{4}\, g_{ij} \left(
        f_{ks}^{\;\;\; s} - f^s_{\; ks} \right),
\label{Eq:dkij}
\end{equation}
where $\eta$ is a free parameter with the only restriction $\eta\neq 0$.

The evolution system (\ref{Eq:gij},\ref{Eq:Kij},\ref{Eq:fkij}), besides
being symmetric hyperbolic, has the additional feature that
all characteristic speeds (with respect to the normal derivative operator
$\partial_0$) are either $0$ or $\pm 1$, so that all characteristic modes
lie either along the light cone or along the orthogonal to the hypersurfaces
$t = const.$ direction.

The particular case with $\eta = 4$
corresponds to the system derived in \cite{ay}. As we
will show shortly, the latter does, however, not seem to admit a symmetric
hyperbolic formulation for the evolution of the constraints.

\subsection{The evolution of the constraints}

In order to solve Einstein's vacuum equations, one has to supplement the
evolution equations (\ref{Eq:gij},\ref{Eq:Kij},\ref{Eq:fkij}) with the
following constraints:
\begin{equation}
\begin{array}{rll}
0 =& C \equiv \frac{1}{2} g^{ab}\partial^k (d_{abk} - d_{kab}) + l.o.\, , & 
\nonumber\\
0 =& C_j \equiv \partial^a K_{aj} - g^{ab}\partial_j K_{ab} + l.o.\, , & 
\nonumber\\
0 =& C_{kij} \equiv d_{kij} - \partial_k g_{ij}\, , & 
\nonumber\\
0 =& C_{lkij} \equiv \partial_{[l} d_{k]ij}\, . & \
\nonumber
\end{array}
\label{Eq:Constraints}
\end{equation}
where $C,C_i$ are the Hamiltonian and momentum constraints, respectively. 

Using the evolution system (\ref{Eq:gij},\ref{Eq:Kij},\ref{Eq:fkij})
for 
the main variables, one obtains the following principal part for the evolution system for
the constraint variables \cite{kst}:
\begin{eqnarray}
\partial_0 C &=& \frac{\eta}{4}\, \partial^k C_k + l.o.\, , 
\label{Eq:C}\\
\partial_0 C_i &=& \frac{4-2\eta}{\eta}\, \partial_i C 
 - \partial^k C^s_{\; kis} - \partial^k C_{kis}^{\;\;\;\; s} + l.o.\, ,
\label{Eq:Ci}\\
\partial_0 C_{kij} &=& l.o.\, ,
\label{Eq:Ckij}\\
\partial_0 C_{lkij} &=& \eta\, \partial_{[l} g_{k](i} C_{j)} 
 + \frac{\eta-4}{4}\, g_{ij} \partial_{[l} C_{k]} + l.o. \, . 
\label{Eq:Clkij}
\end{eqnarray}
A system is symmetrizable if we can find a transformation that
brings the principal part into symmetric form. In order to investigate
under which conditions this system is symmetrizable
hyperbolic, we split $C_{lkij}$ in its
trace and trace-less parts:
$$
C_{lkij} = E_{lkij} + \frac{1}{2} \left( g_{l(i} B_{j)k} - g_{k(i} B_{j)l} \right) 
 + \frac{1}{3}\, g_{ij} W_{lk}\, ,
$$
where $E_{lkij}$ is trace-less with respect to all pair of indices
and where in terms of the traces $S_{ki} \equiv C^s_{\; (ki)s}$, 
$A_{ki} \equiv C^s_{\; [ki]s}$ and $V_{lk} \equiv C_{lks}^{\;\;\;\; s}$,
$$
B_{ki} = \frac{4}{3}\, S_{ki} - \frac{12}{5}\, A_{ki} - \frac{4}{5}\, V_{ki}\, ,\qquad
W_{lk} = \frac{9}{5}\, V_{lk} + \frac{12}{5}\, A_{lk}\, .
$$
Rewriting the principal part in terms of these variables, we
find
\begin{eqnarray}
\partial_0 C &=& \frac{\eta}{4}\, \partial^k C_k + l.o.\, , 
\label{Eq:Ct}\\
\partial_0 C_i &=& \frac{4-2\eta}{\eta}\, \partial_i C 
 - \partial^k S_{ki} - \partial^k A_{ki} - \partial^k V_{ki} + l.o.\, ,
\label{Eq:Cti}\\
\partial_0 S_{ki} &=& -\frac{3\eta}{4}\, \left( \partial_{(k} C_{i)} -
\frac{1}{3}\, g_{ki} \partial^s C_s \right) + l.o.\, ,
\label{Eq:Ski}\\
\partial_0 A_{ki} &=& (1-\eta)\, \partial_{[k} C_{i]} + l.o.\, ,
\label{Eq:Aki}\\
\partial_0 V_{ki} &=& \left( \frac{7\eta}{4} - 3 \right) \partial_{[k} C_{i]} + l.o.\, ,
\label{Eq:Vki}\\
\partial_0 C_{kij} &=& l.o. \, . 
\label{Eq:Ctkij}\\
\partial_0 E_{lkij} &=& l.o. \, . 
\label{Eq:Elkij}
\end{eqnarray}
Having split all the variables into their trace and trace-less parts
the only natural transformations that remain are rescaling of the
variables or linear combinations of the antisymmetric symbols $A_{ki}$
and $V_{ki}$.  First, looking at the terms depending on $C_i$ in the
evolution equation for $C$ and vice-versa, we see that $\eta < 2$ is
needed in order to make the corresponding block in the principal part
symmetric by a rescaling of $C$.  Next, looking at the terms involving
$S_{ki}$ in the evolution equation for $C_i$ and vice-versa, through a
similar reasoning, we obtain the condition $\eta > 0$.  Finally, if $0
< \eta < 2$, it is easy to see that the transformation
$$
\tilde{A}_{ki} = A_{ki} + V_{ki}\; ,\qquad
\tilde{V}_{ki} = (7\eta/4 - 3) A_{ki} + (\eta-1) V_{ki}\; ,
$$
brings the corresponding block into manifestly symmetric form.

We can summarize the result as follows:
If $0< \eta < 2$, the principal part of the system (\ref{Eq:Ct}-\ref{Eq:Elkij})
is symmetric with respect to the inner product associated with
\begin{equation}
\left( U, U \right) \equiv 
\frac{16-8\eta}{\eta^2}\,  C C +  C^i C_i + 
\frac{4}{3\eta}\,  S^{ki} S_{ki}
 + \frac{4}{8 - 3\eta}\,  \tilde{A}^{ki} \tilde{A}_{ki} + 
 \tilde{V}^{ki} \tilde{V}_{ki}
 + C^{kij} C_{kij} + E^{lkij} E_{lkij}\, ,
\label{Eq:CIP}
\end{equation}
where $U = (C,C_i, S_{ki}, \tilde{A}_{ki}, \tilde{V}_{ki}, E_{lkij})^T$.
It is interesting to notice that in a numerical empirical search
for a  value of $\eta$ that improves the stability of a single black hole
evolution, Kidder, Scheel and Teukolsky found the value $\eta = 4/33$,
 which lies inside the interval $0 < \eta < 2$ \cite{kst}. On the other hand, 
 the evolution of the original EC ($\eta =4$), 
for which we were not able
to find a symmetrizer, according to \cite{kst} does not perform very well in 3D 
black hole evolutions. Also of interest is that in the recent work of
Lindblom and Scheel \cite{LiSch} they note that the previously mentioned
range of $\eta$ is 
also preferred. (See figure 5 of their paper; the range $0<\eta<2$ 
translates to $-\infty <\gamma<-0.5$). In that work they also study 
the dependence on another parameter $\hat{z}$, which corresponds to a
rescaling of a variable and therefore its effects do not influence
the principal part of the equations, which is what determines the
level of hyperbolicity of the system.

\section{Characteristic variables}
\label{cpbc}

Here we discuss the characteristic variables of the main evolution
system (\ref{Eq:gij},\ref{Eq:Kij},\ref{Eq:fkij}) and of the evolution
system (\ref{Eq:Ct}-\ref{Eq:Elkij}) for the constraints. The
characteristic variables are needed in order to give the boundary
conditions of type (\ref{Eq:MaxDiss}), which yield a well-posed
initial-boundary value problem.

{}From now on we will concentrate on linearized (around Minkowski
spacetime in Cartesian coordinates) gravity 
for simplicity. That is, the background metric is
$$
ds^2 = -dt^2 + \delta_{ij} dx^i dx^j\, ,
$$
where $\delta_{ij}$ is the flat space metric.
We also assume that our perturbations have vanishing shift
and vanishing linearized densitized lapse.
In these cases, all lower order terms vanish (with the 
obvious exception of the right hand side (RHS) of (\ref{Eq:gij})) and
the evolution equations simplify considerably.

We also choose our spatial domain to be a box
$\Omega = [x_{min},x_{max}] \times [y_{min},y_{max}] \times
[z_{min},z_{max}]$, though the analysis
below could be generalized to other domains.

\subsection{Characteristic variables for the main system}

When linearized around flat spacetime, the main evolution
system reduces to
\begin{eqnarray}
\partial_t g_{ij} &=& -2 K_{ij} \; ,\nonumber\\
\partial_t K_{ij} &=& -\partial^k f_{kij}\; ,\label{kdot}\\
\partial_t f_{kij} &=& -\partial_k K_{ij}\; .\label{ddot}
\end{eqnarray}
Since $g_{ij}$ does not appear in the evolution equations
for $K_{ij}$ and $f_{kij}$, we do not consider its evolution
equation in the following. 
Therefore, the system we consider has the simple form 
$\partial_t u = A^j\partial_j u$, with $u = (K_{ij}, f_{kij})^T$.

The characteristic variables with respect to a direction $n^i$
are variables with respect to which the symbol
$P(\vec{n}) = A^j n_j$ is diagonal. They can be obtained by first
finding a complete set,  $e_1,...,e_{24}$, of eigenvectors
of the symbol (the $e_j$'s are called characteristic modes)
and then expanding $u$ with respect to these vectors. The
coefficients in this expansion are called the characteristic
variables.

For the above system there are six characteristic variables with speed $1$, 
six with speed $-1$ and twelve with zero speed. They are given by
\begin{equation}
\begin{array}{rll}
v_{ij}^{(+)} & = K_{ij} - f_{nij}\; , &\hbox{(speed $+1$)}\\
v_{ij}^{(-)} & = K_{ij} + f_{nij}\; , &\hbox{(speed $-1$)}\\
v_{kij}^{(0)} & = f_{kij} - n_k f_{nij}\; ,&\hbox{(speed $0$)}
\end{array}
\label{Eq:DefCVMain}
\end{equation}
where $f_{nij} \equiv f_{kij} n^k$.
In terms of these variables, the evolution system becomes
\begin{eqnarray}
\partial_t v_{ij}^{(\pm)} &=& \pm \partial_n v_{ij}^{(\pm)} - 
\delta^{kl}\partial_k^T v_{lij}^{(0)}\; ,
\label{Eq:vij}\\
\partial_t v_{kij}^{(0)} &=& -\frac{1}{2}\partial_k^T 
\left( v_{ij}^{(+)} + v_{ij}^{(-)} \right),
\label{Eq:vkij}
\end{eqnarray}
where $\partial_n$ denotes the derivative in the direction of $n$ and
$\partial_k^T \equiv \partial_k - n_k\partial_n$ are the derivatives
with respect to the directions that are orthogonal to $n$.

\subsubsection{Gauge, physical, and constraint-violating modes}
Before we proceed, it is interesting to give the following interpretation
to the characteristic variables of the main system with respect to a fixed
direction $n^k$:
Consider a Fourier mode of $u$ with wave-vector along $n^k$, i.e.
assume that the spatial dependence of $u$ has the form $e^{i n_j x^j}$.
In this case, the linearized constraints assume the form
$L(n^i)u = 0$, where $L(n^i)$ is a constant matrix. Also, since in the
case we are
considering the non principal terms vanish, characteristic modes of
this form solve the evolution equations.

Now we can check which characteristic modes (or combination thereof)
satisfy the constraint equations and which do not. It turns out that all
modes violate the constraints, except for the ones corresponding 
to the characteristic variables $v_{nn}^{(\pm)}$ and 
$\hat{v}_{AB}^{(\pm)} \equiv v_{AB}^{(\pm)} - \delta_{AB} v_{\; C}^{C\, (\pm)}/2$,
where $A,B$ denote directions which are orthogonal to $n$.

Next, consider an infinitesimal coordinate transformation
$x^\mu \mapsto x^\mu + X^\mu$. With respect to it,
\begin{eqnarray}
K_{ij} &\mapsto& K_{ij} - \partial_i\partial_j X^t\; ,\nonumber\\
g_{ij} &\mapsto& g_{ij} + 2\partial_{(i} X_{j)}\; .\nonumber
\end{eqnarray}
Assuming that $X^t$ and $X_j$ are proportional to $e^{i n_j x^j}$
and using (\ref{Eq:dkij}) we find that
$$
v_{nn}^{(\pm)} \mapsto v_{nn}^{(\pm)} + X^t \pm X_n\; ,\qquad
v_{Aij}^{(0)} \mapsto v_{Aij}^{(0)} - \delta_{A(i}\left( n_{j)} X_n - X_{j)} \right) 
 + \frac{\eta-4}{4\eta}\, \delta_{ij} X_A\; ,
$$
while the remaining variables are invariant.
In particular, we can gauge away the variables $v_{nn}^{(+)}$ and
$v_{nn}^{(-)}$. The characteristic variables $\hat{v}_{AB}^{(\pm)}$
are gauge-invariant and satisfy the constraints.

This suggests the following classification: We call
\begin{itemize}
\item the variables $v_{nn}^{(\pm)}$ {\em gauge variables.} 
\item the
variables $\hat{v}_{AB}^{(\pm)}$ {\em physical variables.} 
\item and the 
remaining variables $v_{\; B}^{B\, (\pm)}$ and $v_{nB}^{(\pm)}$
{\em constraint violating variables.}
\end{itemize}
We stress that this classification is exact only if all the fields
depend on the variable $n_k x^k$ and the time coordinate $t$ only. 
(In particular, it is exact in $1+1$ dimensions since then there is
only one space dimension.) Nevertheless, this classification sheds 
light on the boundary treatment below: 
For example, we will see that giving boundary data that is consistent
with the constraints will fix a combination of the in- and outgoing
constraint violating variables while we will be free to
choose any data for some combination of the in- and outgoing gauge and
physical variables (see equation (\ref{Eq:MaxDissCPBC}) below).

\subsection{Characteristic variables for the constraints}

For our purpose, it is sufficient to find the constraint
characteristic variables that have non-zero speed, since these are
the ones that enter the boundary condition (\ref{Eq:MaxDiss}). 
In the system considered here 
there are three characteristic variables with speed $1$ and three with speed $-1$,
\begin{equation}
V_i^{(\pm)} = -C_i \pm \frac{2\eta - 4}{\eta}\, C n_i \pm (S_{ni} + \tilde{A}_{ni}),
\label{Eq:DefCVCons}
\end{equation}
while the remaining variables have zero speed.

\section{Constraint-preserving boundary conditions}
\label{cpbcs}

Here we start with boundary conditions for the constraints that ensure that they 
are preserved though evolution and then translate them into the boundary 
conditions for the main system. In order to do so we write the in- and outgoing characteristic
variables of the constraints in terms of (derivatives of) the characteristic variables
of the main system. Well-posedness of the resulting constraint system
is shown in the Section \ref{well_posedness}.

In order to preserve the constraints we impose boundary conditions for
their evolution of
the form (\ref{Eq:MaxDiss}) with $g=0$ ($g\neq 0$ would not preserve
the constraints). That is, 
\begin{equation}
V_i^{(+)} = L_i^{\; j} V_j^{(-)},
\label{Eq:CPBC1}
\end{equation}
where the matrix $L$ is constant.  As discussed in Sect. \ref{energy}, 
the coupling matrix $L$ must be ``small enough'' if we want to obtain a useful energy
estimate. We will analyze this question in the Section \ref{well_posedness}

Next, our task is to translate the conditions (\ref{Eq:CPBC1})
into conditions on the main variables. Using the definition
of the constraints, Eqs. (\ref{Eq:Constraints}), we find
\begin{eqnarray}
C &=& \frac{\eta}{4}\; \partial^k v_k\; ,\label{cons1}\\
C_i &=& \partial^j K_{ij} - \partial_i K\; ,\label{cons2}\\
S_{ni} + \tilde{A}_{ni} &=&
\partial^l f_{nil} - \partial_if^s_{\;\; sn}  + \frac{\eta}{4}\, 
\left[ n_i\partial^l v_l - 3\partial_i v_n \right], \label{cons3}
\end{eqnarray}
where $v_k = f_{ks}^{\;\;\;\; s} - f^s_{\;\; sk}$.
Using Eqs. (\ref{cons1},\ref{cons2},\ref{cons3}) and the definition of the characteristic variables,
Eqs. (\ref{Eq:DefCVMain},\ref{Eq:DefCVCons}), one obtains
\begin{eqnarray}
V_n^{(\pm)} &=& \partial_n v_{BB}^{(\pm)} - \partial^B v_{Bn}^{(\pm)}
 \pm\frac{1}{2}\left( 1 - \frac{3\eta}{4} \right)\partial^B\left( 2v_{kkB}^{(0)} - 2v_{Bkk}^{(0)}
 - v_{nB}^{(+)} + v_{nB}^{(-)} \right), 
\label{normal1}\\
V_A^{(\pm)} &=& -\partial_n v_{nA}^{(\pm)} - \partial^B v_{AB}^{(\pm)} + \partial_A v_{kk}^{(\pm)}
 \mp\frac{1}{2}\left( 1 - \frac{3\eta}{4} \right)\partial_A
\left( 2v_{kkn}^{(0)} + v_{BB}^{(+)} - v_{BB}^{(-)} \right)
 \label{normal2} \; ,
\end{eqnarray} 
where the capital indices $A,B=1,2$ refer to transverse directions (i.e.
directions orthogonal to $n$), and where we sum over equal indices.

A first problem that arises in the above expressions is that
with maximally dissipative boundary conditions one does not 
control normal derivatives $\partial_n$ of the incoming characteristic
variables at the boundary and, therefore, cannot impose the above conditions,
Eqs. (\ref{normal1},\ref{normal2}).  
However, since in Eqs. (\ref{normal1},\ref{normal2}) the normal
derivatives are present only on
variables with speed $\pm 1$, we can use the equations (\ref{Eq:vij})
 to trade normal derivatives by time and transverse derivatives.
Doing so one obtains
\begin{eqnarray}
V_n^{(\pm)} &=& \pm\partial_t v_{BB}^{(\pm)} - 
\partial^B v_{Bn}^{(\pm)} \pm \partial^C v_{CBB}^{(0)}
 \pm\frac{1}{2}\left( 1 - \frac{3\eta}{4} \right)\partial^B\left( 
2v_{kkB}^{(0)} - 2v_{Bkk}^{(0)}
 - v_{nB}^{(+)} + v_{nB}^{(-)} \right), 
\label{Eq:Vn}\\
V_A^{(\pm)} &=& \mp\partial_t v_{nA}^{(\pm)} - \partial^B
v_{AB}^{(\pm)} + 
\partial_A v_{kk}^{(\pm)}
 \mp \partial^C v_{CnA}^{(0)}
 \mp\frac{1}{2}\left( 1 - \frac{3\eta}{4} 
\right)\partial_A\left( 2v_{kkn}^{(0)} + v_{BB}^{(+)} - v_{BB}^{(-)} \right)
\label{Eq:VA} \; .
\end{eqnarray} 

The second problem is to show well-posedness for the evolution system
of the main variables. Conditions (\ref{Eq:CPBC1}) together with 
Eqs. (\ref{Eq:Vn},\ref{Eq:VA})
do not directly translate into conditions of the form (\ref{Eq:MaxDiss})
for the main variables since transverse derivatives 
appear in the expressions (\ref{Eq:Vn},\ref{Eq:VA}) \footnote{Note that this difficulty 
does not arise in $1+1$ dimensions since then there are no transverse
derivatives. In that case one can simply set the ingoing constraint
violating variables $v_{BB}^{(+)}$ and $v_{nA}^{(+)}$ to zero.}.
In order to get a well-posed initial-boundary problem we look for 
appropriate linear combinations $g = u^{(+)} - R u^{(-)}$ of in- and outgoing
main variables and an appropriate coupling matrix $L$ such that the conditions
(\ref{Eq:CPBC1}) with (\ref{Eq:Vn},\ref{Eq:VA})
can be incorporated in a closed set of evolution system
at the boundary for the variables $g$ and some variables with zero speed.
This is discussed next.

\subsection{Neumann boundary conditions: evolution system on each face}
\label{neumann_bc}

Consider the combinations
$d_{ij} = v_{ij}^{(+)} - v_{ij}^{(-)}$ and $s_{ij} = v_{ij}^{(+)} +
v_{ij}^{(-)}$ ($d_{ij}$ stands for {\bf d}ifference \footnote{Notice
that $d$ is used for this quantity as well as for $d_{kij}=\partial _k
g_{ij}$, equation (\ref{Eq:dkij}). There does not seem to be risk of confusion since both
objects have different number of indexes.}, and $s_{ij}$ for
{\bf s}um). 

Noticing that 
\begin{eqnarray*}
V_n^{(+)} + V_n^{(-)} &=& \partial_t d_{BB} - \partial^A s_{An}\; ,
\\
 -V_A^{(+)} + V_A^{(-)} &=& \partial_t s_{An} + \partial^B d_{AB} -
  \partial_A d_{BB} - \partial_A d_{nn} + 2\partial^C v_{CnA}^{(0)} +
  \left( 1 - \frac{3\eta}{4} \right)\partial_A\left( 2v_{BBn}^{(0)} +
  d_{BB} \right),
\end{eqnarray*}
we see that one way of imposing boundary conditions such
that the 
constraints are satisfied is through
\begin{eqnarray}
0 & = & V_n^{(+)} + V_n^{(-)} \label{Eq:bondy1} \; , \\
0 & = & -V_A^{(+)} + V_A^{(-)} \label{Eq:bondy2} \; .
\end{eqnarray}
Equations (\ref{Eq:bondy1},\ref{Eq:bondy2}) amount to giving as boundary conditions
for the constraints' propagation a coupling between the incoming and outgoing
characteristic constraint variables as in Eq. (\ref{Eq:CPBC1}) with
$(L_i^{\; j}) = diag(-1,1,1)$. These 
equations can also be seen as equations for $d_{BB}$ and $s_{nA}$ at
the $n$-face, where 
$\hat{d}_{AB}$ and $d_{nn}$ are a priori prescribed functions
 (i.e. they are not fixed by the constraints' treatment). Here
$\hat{d}_{AB}$ is defined as the traceless part of $d_{AB}$. That is,
\begin{displaymath}
\hat{d}_{AB} = d_{AB} - \frac{1}{2}\delta _{AB} d_C^{\;\;C}
\end{displaymath}
where $\delta _{AB}=1$ if $A=B$ and zero otherwise.

In fact, Eqs. (\ref{Eq:bondy1},\ref{Eq:bondy2}) do not not contain as
dynamical variables only $d_{BB}$ and $s_{nA}$, but also some zero
speed variables. Therefore, in order to get a closed system, we need 
evolution equations for the zero speed variables
$v_{ABn}^{(0)}$. These equations can be obtained from the evolution 
system of the main variables, equation (\ref{Eq:vkij}):
\begin{equation}
0 = \partial_t v_{ABn}^{(0)} + \frac{1}{2}\,\partial_A s_{nB}\; .
\label{Eq:bondy3}
\end{equation}
Below we will show that equations
(\ref{Eq:bondy1},\ref{Eq:bondy2},\ref{Eq:bondy3}) constitute a
symmetrizable hyperbolic system of evolution equations at the face
orthogonal to $n$ for the variables $(d_{BB}, s_{An}, v_{ABn}^{(0)})$, 
where $\hat{d}_{AB} = d_{AB} - 1/2\, \delta_{AB} d_{CC}$ and $d_{nn}$
can be freely prescribed. Since the domain we are interested in is a
box, Eqs. (\ref{Eq:bondy1},\ref{Eq:bondy2},\ref{Eq:bondy3}) need
to be evolved at each face. In order to do so one needs boundary
conditions at each edge (intersection of two faces). Below we will
also explain how these boundary conditions are naturally fixed by
compatibility conditions.

Assuming one already has the solution to
Eqs. (\ref{Eq:bondy1},\ref{Eq:bondy2},\ref{Eq:bondy3}) at each face,
the boundary conditions for the main variables, which are of the
required form (\ref{Eq:MaxDiss}), are the following:
\begin{equation}
v_{BB}^{(+)} = v_{BB}^{(-)} + d_{BB}\; ,\qquad
v_{nA}^{(+)} = -v_{nA}^{(-)} + s_{nA}\; ,\qquad
v_{nn}^{(+)} = v_{nn}^{(-)} + d_{nn}\; ,\qquad
\hat{v}_{AB}^{(+)} = \hat{v}_{AB}^{(-)} + \hat{d}_{AB}\; ,
\label{Eq:MaxDissCPBC}
\end{equation}
where $d_{BB}$ and $s_{nA}$ are obtained from the evolution system
(\ref{Eq:bondy1}-\ref{Eq:bondy3}) at the $n$-face and where
we can specify the gauge variable $d_{nn}$ and the physical variable
$\hat{d}_{AB}$ freely. Note that
\begin{equation}
d_{nn} = -2f_{nnn}\; ,\qquad
\hat{d}_{AB} = -2f_{nAB} + \delta_{AB} f_{nCC}\; .   \label{neumann_data}
\end{equation}
In view of equation (\ref{Eq:dkij}) and the fact that if the
constraints are satisfied we have $d_{kij} = \partial_k g_{ij}$,  
we see that these are Neumann conditions on some
components of the three-metric.

Now we go back to the evolution system
(\ref{Eq:bondy1}-\ref{Eq:bondy3}) at the face and look at it 
in detail. In order to make
the notation more compact we write $d^{(n)} =
d_{BB}$, $s_A^{(n)} = s_{An}$, $h_{AB}^{(n)} = v_{ABn}^{(0)}$ for the
variables associated with the face orthogonal to $n$. Then we have
\begin{eqnarray}
\partial_t d^{(n)} &=& \partial^A s_A^{(n)}\; ,\label{neumann1}\\
\partial_t s_A ^{(n)} &=& -2\partial^B h_{BA}^{(n)} - \frac{4-3\eta}{2}\, 
\partial_A h^{(n)} - \frac{2-3\eta}{4}\, \partial_A d^{(n)}
  - \partial^B \hat{d}_{AB}^{(n)} + \partial_A d_{nn}^{(n)}, \label{neumann2}\\
\partial_t h_{AB}^{(n)} &=& -\frac{1}{2}\, \partial_A s_B^{(n)}\; , \label{neumann3}
\end{eqnarray}
where $h^{(n)}$ denotes the trace of $h_{AB}^{(n)}$.  
One can check that as long as $\eta\neq 2$ this system is symmetric hyperbolic
with respect to the inner product associated with
\begin{equation}
(B,B) \equiv \left(d^{(n)}\right)^2 + 2 s_A^{(n)} s_B^{(n)} \delta ^{AB} + 
\left[ (6-3\eta)\, h^{(n)}_{AB}\delta^{AB} + 
\frac{4 - 3\eta}{2}\, d^{(n)} \right]^2
  + 8\hat{h}^{(n)}_{AB} \hat{h}^{(n)}_{CD} \delta ^{AC} \delta ^{BD}\; ,
\label{B,B}
\end{equation}
where $B = (d,s_A,h_{AB})$ and $\hat{h}_{AB}$ denotes the
trace-less part of $h_{AB}$, $\hat{h}_{AB} = h_{AB} - (\delta _{AB} h_C^{\;\;C})/2$.

\subsection{
Neumann boundary conditions: Compatibility  conditions at the 
edges of the faces}

The system we just introduced is defined on each of the six $n$-faces. The
faces themselves have boundaries: the edges that join them. We need to
ensure that the system on each face is well posed taking into account
the boundary conditions at the edges.  Since each edge is shared by
two faces, this will translate into compatibility conditions among the
various systems.

The system of interest has, at each edge, two ingoing modes with speeds
$\sqrt{3/2}$ and $1$ and two outgoing modes with speeds
$-\sqrt{3/2}$ and $-1$. At face $n$, the corresponding characteristic
variables with respect to a direction $m^A$, are
\begin{eqnarray}
w_{\pm\sqrt{3/2}} &=& \pm\frac{1}{2}\sqrt{\frac{2}{3}}\, s^{(n)}_m
 - \frac{1}{3}\left[ 2h^{(n)}_{mm} + \frac{4-3\eta}{2}\, h^{(n)} + 
\frac{2-3\eta}{4}\, d^{(n)} \right],
\nonumber\\
w_{\pm 1} &=& \pm s_p^{(n)} - 2h^{(n)}_{mp}\; ,
\nonumber
\end{eqnarray}
where $p$ is a transverse direction orthogonal to $m$ (and $n$).

In terms of the original variables, the variables $d^{(n)}, s_A^{(n)},
h_{AB}^{(n)}, d_{nn}^{(n)}, \hat{d}_{AB}^{(n)}$ on the $n$-face are
$$
d^{(n)} = -2f_{nBB}\; ,\qquad
s_A^{(n)} = 2K_{nA}\; ,\qquad
h_{AB}^{(n)} = f_{ABn}\; ,\qquad
d_{nn}^{(n)} = -2f_{nnn}\; ,\qquad
\hat{d}^{(n)}_{AB} = -2f_{nAB} + \delta_{AB} f_{nCC}\; .
$$
Now let $\{n,m,p\}$ be orthogonal directions naturally associated with
the box (say, the standard Cartesian coordinates $\{x,y,z\}$).
If we compare these variables in two different directions $n$ and $m$,
say, we obtain the following compatibility conditions:
\begin{eqnarray}
s_m^{(n)} &=& s_n^{(m)}\; ,\label{comp1}\\
 -2h_{mp}^{(n)} &=& \hat{d}_{pn}^{(m)}\; ,\label{comp2}\\
h_{pm}^{(n)} &=& h_{pn}^{(m)}\; . \label{comp3}
\end{eqnarray}
On the other hand, we have, by definition of the characteristic variables,
$$
w_{+\sqrt{3/2}}^{(n)} - w_{-\sqrt{3/2}}^{(n)} = \sqrt{\frac{2}{3}}\, s_m^{(n)}\; ,\qquad
w_{+1}^{(n)} + w_{-1}^{(n)} = -4h_{mp}^{(n)}\; .
$$
Therefore, the correct boundary conditions at the edges are
\begin{eqnarray}
w_{+\sqrt{3/2}}^{(n)} &=& w_{-\sqrt{3/2}}^{(n)} + \sqrt{\frac{2}{3}}\,
s_n^{(m)}\; ,
\label{boundary_neumann1}\\
w_{+1}^{(n)} &=& -w_{-1}^{(n)} + 2\hat{d}_{np}^{(m)}\; . \label{boundary_neumann2}
\end{eqnarray}
Notice that up to now the quantities that were freely specified were
$d_{nn}$ and $\hat{d}_{AB}$ one each of the $n$-faces. Therefore, in order to fix
the boundary data for the systems that are defined on these faces we also have to a priori
specify the quantities $s_n^{(m)}=2K_{nm}=s_m^{(n)}$ at the
edges defined by the intersection of faces $n$ and $m$.

Imposing these boundary conditions automatically implies that the
compatibility conditions (\ref{comp1}) and (\ref{comp2}) are
satisfied. On the other hand, we have at the edge joining the faces
$n$ and $m$, 
\begin{displaymath}
\partial_t \left( h_{pm}^{(n)} - h_{pn}^{(m)}\right) = -\frac{1}{2}\partial_p
\left(s_m^{(n)}- s_n^{(m)}  \right) = 0
\end{displaymath}
where the last equality follows from imposing the boundary conditions
at the edges. Therefore, these boundary conditions also imply that the compatibility
condition (\ref{comp3}) is satisfied through evolution provided it
does so initially.

\subsection{Dirichlet boundary conditions}
\label{dirichlet_bc}

In a similar way to the Neumann case, one can obtain a closed system at the boundary
for the variables $(s_{BB}, d_{An}, v_{Ann}^{(0)}, v_{ABB}^{(0)},
v_{BBA}^{(0)})$
by requiring
\begin{eqnarray}
0 &=& V_n^{(+)} - V_n^{(-)} \; , \label{Eq:bondy4} \\
0 &=& V_A^{(+)} + V_A^{(-)} \; , \label{Eq:bondy5}
\end{eqnarray}
where 
\begin{eqnarray*}
V_n^{(+)} - V_n^{(-)} &=& \partial_t s_{BB} - \partial^A d_{An} +
  2\partial^A v_{ABB}^{(0)} + \left( 1 - \frac{3\eta}{4}
  \right)\partial_A\left( 2v_{BBA}^{(0)} - 2v_{Akk}^{(0)} - d_{nA}
  \right) \; ,\\
 -V_A^{(+)} - V_A^{(-)} &=& \partial_t d_{An} + \partial^B s_{AB} -
\partial_A s_{BB} - \partial_A s_{nn}\;. 
\end{eqnarray*}
One also has to take into account the evolution equations for the zero speed variables:
\begin{equation}
0 = \partial_t v_{Aij}^{(0)} + \frac{1}{2}\, \partial_A s_{ij}\; .
\label{Eq:bondy6}
\end{equation}
In this case, one can freely specify $s_{nn}$ and $\hat{s}_{AB}$, which 
corresponds to Dirichlet conditions on some components of the
extrinsic curvature:
\begin{equation}
s_{nn} = 2K_{nn} \;\;\; , \;\;\; 
\hat{s}_{AB} = 2K_{AB} - \delta_{AB} K_{CC}\; . \label{dirichlet_cond}
\end{equation}

Assuming one already has the solution to
Eqs. (\ref{Eq:bondy4},\ref{Eq:bondy5}) at each face,
the boundary conditions for the main variables, which are of the
required form (\ref{Eq:MaxDiss}), are the following:
\begin{equation}
v_{BB}^{(+)} = -v_{BB}^{(-)} + s_{BB}\; ,\qquad
v_{An}^{(+)} = v_{An}^{(-)} + d_{An}\; ,\qquad
v_{nn}^{(+)} = -v_{nn}^{(-)} + s_{nn}\; ,\qquad
\hat{v}_{AB}^{(+)} = -\hat{v}_{AB}^{(-)} + \hat{s}_{AB}\; .
\label{Eq:MaxDissCPBC_dir}
\end{equation}

In terms of the variables $s = s_{BB}$, $d_A = d_{An}$,
$h_A = 3\eta\, v_{ABB}^{(0)} + (4-3\eta)(v_{BBA}^{(0)} - v_{Ann}^{(0)})$,
we can rewrite the boundary system 
(\ref{Eq:bondy4}-\ref{Eq:bondy6}) as
\begin{eqnarray}
\partial_t s &=& \frac{8-3\eta}{4}\,\partial^A d_A -
\frac{1}{2}\,\partial^A h_A\; ,\label{dir1}
\\ \partial_t d_A &=&
\frac{1}{2}\,\partial_A s + \partial_A s_{nn} -
\partial^B\hat{s}_{BA}\; , \label{dir2}
\\ \partial_t h_A &=&
-\frac{4+3\eta}{4}\, \partial_A s +
\frac{4-3\eta}{2}\,\left(\partial_A s_{nn} - \partial^B
\hat{s}_{BA}\right).  \label{dir3}
\end{eqnarray}
This system is symmetrizable hyperbolic with respect to the inner
product associated with
\begin{equation}
(B,B) =  4s^2 + 8d^A d_A + \left(
 h^A - \frac{4-3\eta}{2}\, d^A \right)\left( h_A - \frac{4-3\eta}{2}\,
 d_A \right) \label{energy_dir}
\end{equation}
where $B = (s,d_A,h_A)$.

There is one ingoing  
and one outgoing mode, with speeds $\pm\sqrt{3/2}$, respectively. 
The corresponding variables in a direction $m^A$ are
$$
w_{\pm\sqrt{3/2}} = \sqrt{3/2}\, s \pm \frac{1}{2}\left(
\frac{8-3\eta}{2}\, d_m - h_m \right).
$$
We now turn to the compatibility at the edges of the faces. 
In terms of the original variables one has
\begin{eqnarray*}
&& s^{(n)} = 2K_{BB}\; ,\qquad
   d_A^{(n)} = -2f_{nnA}\; ,\qquad
   h_{A}^{(n)} = 3\eta\, f_{ABB} + (4-3\eta)(f_{BBA} - f_{Ann}) ,
\nonumber\\
&& s_{nn}^{(n)} = 2K_{nn}\; ,\qquad
   \hat{s}^{(n)}_{AB} = 2K_{AB} - \delta_{AB} K_{CC}\; .
\end{eqnarray*}
{}From these 
expressions one can see that the only  compatibility conditions are
\begin{eqnarray}
s^{(n)} & = & 2\left( s_{mm}^{(m)} - \hat{s}^{(n)}_{mm} \right) \; , 
\label{compatibility_dir1} \\
s^{(m)}_{mm} + 2 \hat{s}^{(m)}_{nn} & = & s^{(n)}_{nn} + 2
\hat{s}^{(n)}_{mm} \; .  \label{compatibility_dir2}
\end{eqnarray}
Equation (\ref{compatibility_dir1}) fixes the boundary data for the
ingoing variable:
\begin{equation}
w_{+\sqrt{3/2}}^{(n)} = -w_{-\sqrt{3/2}}^{(n)} + 4\sqrt{3/2}(
s_{nn}^{(m)} - \hat{s}_{mm}^{(n)}).   \label{boundary_dirichlet}
\end{equation}
This condition will be used in the energy estimate (\ref{Eq:estimate3})
in order to prove well-posedness for the system at
each face in the Dirichlet case. On the other hand, 
Eq. (\ref{compatibility_dir2}) is a compatibility
condition at the intersection of faces $n$ and $m$ for the free
boundary data. 

\section{Well posedness}
\label{well_posedness}

We start by showing that the conditions (\ref{Eq:bondy1},\ref{Eq:bondy2})
and (\ref{Eq:bondy4},\ref{Eq:bondy5}), for the Neumann and Dirichlet
case, respectively, 
imply that the evolution system for the constraints is well-posed.
Then we derive energy estimates for the closed system of evolution equations
at each face, and using these estimates we show that the evolution of the main
system is well-posed as well.

Since we have already shown that all the evolution equations involved
in our constraint-preserving treatment are symmetric hyperbolic, and
since we have already cast all boundary conditions in maximally
dissipative form, the main 
purpose of this section is to explicitly show that the different couplings are
``small enough'' with respect to the corresponding symmetrizers, in
the sense 
discussed in Section \ref{energy}.

\subsection{Constraint propagation}

Here we derive an estimate
for the growth of the energy
$$
E_{constraints} = \int_\Omega (U,U)\, d^3 x,
$$
where $(U,U)$ is defined in (\ref{Eq:CIP}).
Taking a time derivative and using equations (\ref{Eq:Ct}-\ref{Eq:Elkij})
we obtain, after integrations by parts,
\begin{displaymath}
\frac{d}{dt}\, E_{constraints} = 2\int_{\partial\Omega} \left[ 
\frac{4-2\eta}{\eta}\, C C_n - C^i S_{ni} - C^i\tilde{A}_{ni} \right] d\sigma,
\end{displaymath}
where $n^k$ is the unit outward normal to the boundary $\partial\Omega$
of the domain. Expressing the integrand in terms of characteristic
variables defined in (\ref{Eq:DefCVCons}), we obtain
\begin{displaymath}
\frac{d}{dt}\, E _{constraints} = \frac{1}{2}\int_{\partial\Omega} 
\delta^{ij}\left( V_i^{(+)} V_j^{(+)} - V_i^{(-)} V_j^{(-)} \right) d\sigma = 0,
\end{displaymath}
where the last equation follows from the conditions 
(\ref{Eq:bondy1},\ref{Eq:bondy2}) and
(\ref{Eq:bondy4},\ref{Eq:bondy5}) in the Neumann and Dirichlet cases, respectively.
Therefore, the initial-boundary value problem for the constraints
is well-posed. In particular, this implies that zero initial
data for the constraints implies that the constraints are
zero at later times as well.

\subsection{Face systems}

\subsubsection{Neumann conditions}

In order to show well-posedness for each system defined on the $n$-face
$\Gamma _n $ we consider
the corresponding energy norm for the Neumann case,
$$
 E _{(\Gamma _n,N)} = \int_{\Gamma _n} (B,B)\, d\sigma,
$$
where $(B,B)$ is 
defined in (\ref{B,B}).
Taking a time derivative and using the above evolution equations
we obtain
\begin{equation}
\frac{d}{dt} E _{(\Gamma _n,N)} = \int_{\partial\Gamma _n} \left[
  6\sqrt{\frac{3}{2}}\left( w_{+\sqrt{3/2}}^2 - w_{-\sqrt{3/2}}^2 \right) 
   + 4\left( w_{+1}^2 - w_{-1}^2 \right) \right]\, ds
 + 4\int_{\Gamma _n } s^A\left( \partial_A d_{nn} - \partial^B\hat{d}_{AB} \right)\, d\sigma.
\label{Eq:estimate1}
\end{equation}
We use the boundary conditions 
(\ref{boundary_neumann1},\ref{boundary_neumann2}), with $s_n^{(m)}=0$
and $\hat{d}_{np}^{(m)}=0$ for the moment, to get rid of the first term
on the RHS of Eq. (\ref{Eq:estimate1}).  Using Schwarz's inequality, 
the second term is estimated as follows:
$$
4\int_{\Gamma _n } s^A\left( \partial_A d_{nn} - \partial^B\hat{d}_{AB} \right)\, d\sigma
\leq 4\, f\, E _{(\Gamma _n,N)}^{1/2}\; ,
$$
where
$$
f^2 = \int_{\Gamma _n} \left( \partial_A d_{nn} - \partial^B\hat{d}_{AB} \right)
\left( \partial^A d_{nn} - \partial_C\hat{d}^{AC} \right)\, d\sigma.
$$
Therefore, we end up with the estimate
\begin{equation}
E _{(\Gamma _n,N)}(t)^{1/2} \leq  E _{(\Gamma _n,N)}(0)^{1/2} + 2\int_0^t f(s)\, ds
\label{Eq:estimate2}
\end{equation}
and see that the energy is bounded and the bound is determined by 
the norm  $f$ of the free data on the boundary.

The assumptions $s_n^{(m)}=0=\hat{d}_{np}^{(m)}$ can 
be easily relaxed with an argument similar to the one we presented in
the paragraph following equation (\ref{Eq:MaxDiss}). That is, one
defines a new variable that incorporates the inhomogeneity in the
energy generated by the term that appears in the case of a non-smooth
boundary and obtains an energy estimate for the new variable, since
the variable redefinition is finite and well defined. Well
posedness follows immediately.

\subsubsection{Dirichlet conditions}

The Dirichlet case is similar to the Neumann one. The energy is
now given by
$$
 E _{(\Gamma _n,D)} = \int_{\Gamma _n} (B,B)\, d\sigma,
$$
with $(B,B)$ given by Eq. (\ref{energy_dir}). We obtain the estimate
\begin{equation}
\frac{d}{dt} E_{(\Gamma _n, D)} \leq
\sqrt{8/3}\,\int_{\partial\Gamma _n} \left( w_{+\sqrt{3/2}}^2 -
w_{-\sqrt{3/2}}^2 \right) ds + 2\, \tilde{f} E^{1/2}_{(\Gamma _n,D)},
\label{Eq:estimate3}
\end{equation}
where 
$$
\tilde{f}^2 = \int_{\Gamma _n} \left( \partial_A s_{nn} - \partial^B\hat{s}_{AB} \right)
\left( \partial^A s_{nn} - \partial_C\hat{s}^{AC} \right)\, d\sigma,
$$
and we can proceed as in the Neumann case, using in this case the boundary
condition (\ref{boundary_dirichlet}).

\subsection{Main system}

Having obtained a bound for each closed system defined on a face we can
obtain a bound for the main evolution variables by the standard
techniques described in section II where we now have the boundary
conditions (\ref{Eq:MaxDissCPBC}) and (\ref{Eq:MaxDissCPBC_dir}) for
the Neumann and Dirichlet cases, respectively,  which are of the required form
(\ref{Eq:MaxDiss}).

\section{Summary}
\label{summary}

The well posed, constraint preserving boundary
conditions presented in this paper can be summarized as follows. 

\subsection{Neumann case}

\begin{itemize}
\item Free data:

At each face (say, the $n$-one), the three quantities $d_{nn}$ and $\hat{d}_{AB}$
  must be a priori defined (subject to
 standard compatibility conditions with the initial data). These quantities
correspond to the normal (that is, in the $n$ direction) derivative of
the normal and the transverse,
traceless parts of the three metric, see Eq. (\ref{neumann_data}). One
  of these variables is gauge and the other two are physical, in the sense
  discussed in Section \ref{cpbc}. In addition, the quantities
$s_m^{(n)}=2K_{nm}$ should   be prescribed at each edge defined by the
intersection of faces $n$ and $m$. These quantities must satisfy the 
 compatibility conditions $s_n^{(m)}=s_m^{(n)}$. 

\item Evolution systems on faces:

The $2D$ symmetric hyperbolic $7\times 7$ system
(\ref{neumann1},\ref{neumann2},\ref{neumann3}) 
is evolved on each face. This system needs boundary conditions at
the edges of the corresponding face. They are given by equations
(\ref{boundary_neumann1}, \ref{boundary_neumann2}). 

The solution to each of these systems 
provides the three quantities $d^{(n)}_{BB}$ and
$s^{(n)}_{nA}$ at each of the six $n$-faces.

\item Main evolution system:

The main system (\ref{kdot},\ref{ddot}) is evolved in the $3D$ domain. This
system needs boundary conditions, at each face, for the six incoming
characteristic modes. These boundary conditions at, say face $n$, are given by
 Eq. (\ref{Eq:MaxDissCPBC}), where the needed information for three of these
boundary conditions 
is provided by the a priori specified $d_{nn}$ and $\hat{d}_{AB}$, while the other three are
given by  $d^{(n)}_{BB}$ and $s^{(n)}_{nA}$. 

\end{itemize}

\subsection{Dirichlet case}

\begin{itemize}

\item Free data:

At each face (say, the $n$ one), now the three quantities $s_{nn}$ and $\hat{s}_{AB}$
 must be a priori given. They correspond to the time
derivative of the normal and transversal, traceless part of the three
metric, see Eq. (\ref{dirichlet_cond}). These three quantities have to
 satisfy the standard compatibility
 conditions  with the initial data, but also some compatibility
 conditions at 
 edges, Eqs. (\ref{compatibility_dir1},\ref{compatibility_dir2}). They
 are also gauge and physical variables, in the sense discussed in Section
 \ref{cpbc}. 

\item Evolution systems on faces:

The $2D$ symmetric hyperbolic $5\times 5$ system
(\ref{dir1},\ref{dir2},\ref{dir3}) 
is evolved on each face. This system needs boundary conditions
at each edge. They are given by
Eq. (\ref{boundary_dirichlet}). 

The solution to each of these systems 
provides the three quantities $s^{(n)}_{BB}$ and
$d^{(n)}_{An}$ at each of the six $n$-faces.

\item Main evolution system:

The main system (\ref{kdot},\ref{ddot}) is evolved in the $3D$ domain. This
system needs boundary conditions, at each face, for the six incoming
characteristic modes. These boundary conditions are given by
(\ref{Eq:MaxDissCPBC_dir}), where the needed information in three of these
boundary conditions 
is provided by the a priori specified $s_{nn}$ and $\hat{s}_{AB}$, while the other three are
given by  $s^{(n)}_{BB}$ and $d^{(n)}_{An}$.
 
\end{itemize}

\section{Conclusions:}
\label{conclusions}

We have studied the system of evolution equations for the constraints
for a subfamily of the generalized Einstein-Christoffel symmetric
hyperbolic system. We have shown how to give boundary data for the
constraints in such a way that it translates into boundary data for
the main system that yields a well posed problem, both for the main
system and the system of evolution equations for the constraints. We
have studied the case of a boundary that is not smooth, as is the case
of the usual cubic boxes used in numerical relativity. This required
additional care at the boundaries of each face, with the ensuing
compatibility conditions. It should be noted that the energy estimates
derived do not necessarily guarantee the existence of a smooth
solution in the presence of corners even with the compatibility
conditions we presented. Further work is needed to establish
smoothness of the solution.

Our analysis was
carried out for the case of linearized gravity around Minkowski
space-time. It is expected that similar techniques will be useful in
the case of other background space-times and also in the non-linear
case. We will discuss these generalizations in future papers. Also,
since we have followed a systematic approach and have not taken any 
advantage of the gauge choice, in principle it should be possible to apply the same
procedure to symmetric hyperbolic formulations with live gauges \cite{st}.

We have also found that, at least with the formulation of Einstein's
equations here used (the generalized EC), the Neumann and Dirichlet
cases are in fact the {\em only} ones for which well posedness can be
established through the techniques used in this paper (see the appendix).
More specifically, we have found that these two cases are the only ones in
which closed systems at the faces can be obtained. However, this does
not mean that these are the only well posed cases, since in the
initial-boundary value problem an energy estimate is a sufficient but
not necessary condition for well posedness. Also, giving the
appropriate $(d_{nn},\hat{d^{(n)}}_{AB},s_{nm})$ data in the Neumann
case, or the appropriate $(s_{nn},\hat{s}_{AB})$ in the Dirichlet
case, one should be able to recover any spacetime in any slicing. This
is because our constraint-preserving treatment makes sure that one is
solving not only Einstein's evolution equations but also the
constraints. But it does not make any restriction on the space of
solutions to the Einstein equations. However, it is not clear how to
choose these ``appropriate'' boundary conditions, without any a priori
knowledge of the solutions, in order to model an isolated source,
given that the boundaries are at a finite distance. This same problem
appears in similar approaches \cite{szilagyi1,szilagyi2}. One possible
solution is to provide these functions through Cauchy-characteristic
\cite{c_match,szilagyi2} or Cauchy-perturbative matching
\cite{p_match}, or to resort arguments using the peeling
property. Another possibility would be to impose a ``no incoming
radiation'' condition. However, it is not clear how to do this within
formulations that have as extra variables first but not second spatial
derivatives of the three metric.  This might be remedied by repeating
this construction for other systems of evolution equations where the
variables that represent gravitational radiation at the boundary play
a more central role, as in \cite{fn}. For instance, a system of
evolution equations of higher order where the Weyl tensor is the
fundamental variable, would be suitable for this purpose.  

With the results of this paper one can now assure that both the
evolution equation for the main variables of the problem and the
evolution equations for the constraints are well posed on a manifold
with (non smooth) boundary. This allows to evolve initial data
that satisfy the constraints beyond their domain of dependence, as is
of interest in numerical simulations of the binary black hole
problem. Moreover, well posedness opens the possibility of
constructing numerical schemes for which numerical stability can be
rigorously proved.

\section{Acknowledgments}
We wish to thank Gabriel Nagy for useful discussions 
 and Alan Rendall for reading the manuscript.
This work was supported in part by grants NSF-PHY-9800973,
NSF-INT-0204937, the Swiss National Science Foundation, by Fundaci\'on
Antorchas and the Horace C. Hearne Jr. Institute of Theoretical
Physics.

\appendix
\section{Closing the boundary system}

In Section \ref{cpbcs} we showed how to construct a closed system at the boundary
by taking appropriate linear combinations of characteristic
variables.  We also chose a particular combination of ingoing and outgoing
constraints and showed that it gives rise to a system of partial
differential equations that lives on the boundary.  To close this
system we had to include the 
evolution of some zero speed modes.  A question that arises is whether there other ways of closing the
boundary system, apart from the Neumann
(\ref{neumann1}-\ref{neumann3}) and the Dirichlet
(\ref{dir1}-\ref{dir3}) case.

To answer this question we will make no assumptions on the coupling
matrices $R$ and $L$.  The 
boundary condition for the system of the constraints is assumed to be
\begin{equation}
V^{(+)}_i = L_i{}^j V^{(-)}_j\,, \label{coupling_constr}
\end{equation}
where $L_i{}^j$ is a $3\times 3$ coupling matrix and $V_i^{(\pm)}$ is
given in (\ref{Eq:Vn}) and (\ref{Eq:VA}). 
At the boundary data must be given to the ingoing modes.  We will
assume that they satisfy
\begin{equation}
v^{(+)}_{ij} = R_{ij}{}^{kl} v^{(-)}_{kl} + b_{ij}\,, \label{coupling_vars}
\end{equation}
where $ R_{ij}{}^{kl}$ is a $6 \times 6$  coupling matrix and $b_{ij}$ is the boundary data.  If we insert
(\ref{coupling_vars}) into (\ref{coupling_constr}) we 
obtain a system which contains
derivatives of the boundary data $b_{ij}$, of the outgoing modes $v^{(-)}_{ij}$,
and of the zero speed modes $v^{(0)}_{Tij}$.
This system can be solved for the time derivatives of three of the
$b_{ij}$, namely $b_{BB}$ and $b_{nA}$.  To the  remaing $b_{ij}$ one
can give arbitrary data and consider them as source terms.
In order to close the system the coefficients that multiply terms
containing outgoing modes $v^{(-)}_{ij}$ must vanish.  After imposing
this condition the coupling matrices $R$ 
and $L$, which in general depend on $45$ parameters, depend on
one parameter only (apart from $\eta$).  The zero speed modes that
appear in the rhs of this system cannot be eliminated by any choice of
this parameter.  Therefore in order to close the system one has to enlarge
it by including the evolution of at least the zero speed modes that
appear in the rhs.  The requirement that the evolution of these zero
speed modes do not contain any spatial derivatives of outgoing modes,
forces the couplings $R$ and $L$ to be the ones the we used in
subsections (\ref{neumann_bc}) and (\ref{dirichlet_bc}).

Summarizing, the Neumann and the Dirichlet cases are the only ways that
one can obtain a closed system at the boundary.  Furthermore, as we
have shown in this paper, the boundary system is symmetric
hyperbolic and the coupling matrices $R$ and $L$ are ``not too large''.
Any other choice of  
coupling matrices would lead 
to a system for which the techniques used in this paper to prove
well-posedness cannot be applied.



\begin{thebibliography}{10}

\bibitem{laxfr} P. D. Lax, R. S. Phillips, Commun. Pure Appl. Math. {\bf 13},
427 (1960).



\bibitem{convergence}
G.~Calabrese, J.~Pullin, O.~Sarbach and M.~Tiglio,
Phys.\ Rev.\ D {\bf 66}, 041501 (2002)


\bibitem{fn} H. Friedrich and G. Nagy, Comm. Math. Phys. {\bf 201}, 619 (1999).

\bibitem{stewart} J. M. Stewart, Class. Quantum Grav. {\bf 15}, 2865
(1998).

\bibitem{iriondo} M. S. Iriondo and O.A. Reula, Phys. Rev. D {\bf 65},
044024 (2002).

\bibitem{szilagyi1} B.Szilagyi, B. Schmidt, and J. Winicour,
Phys. Rev. D {\bf 65}, 
064015 (2002).

\bibitem {bardeen} J.~M.~Bardeen and L.~T.~Buchman,
Phys.\ Rev.\ D {\bf 65}, 064037 (2002).

\bibitem{calabrese} G. Calabrese, L. Lehner, and M. Tiglio, Phys.\ Rev.\ D {\bf 65},
104031 (2002).

\bibitem{szilagyi2} B.~Szilagyi and J.~Winicour,
{\em ``Well-Posed Initial-Boundary Evolution in General Relativity,''}
arXiv:gr-qc/0205044.

\bibitem{kst}
L.E. Kidder, M.A. Scheel, and S.A. Teukolsky,
Phys. Rev. D {\bf 64}, 064017 (2001).

\bibitem{ay}
A. Anderson and J.W. York, Jr., Phys. Rev. Lett. {\bf 82}, 4384 (1999).


\bibitem{kreiss1} H.O. Kreiss, J. Lorenz, {\em ``Initial-Boundary Value
Problems and the Navier-Stokes Equations,''} Academic Press, (1989).

\bibitem{secchi}P. Secchi, Diff. Int. Eq. {\bf 9}, 671 (1996); Arch. Rat. Mech. Anal. 
{\bf 134}, 595 (1996).

\bibitem{rauch} J. Rauch, Trans. Am. Math. Soc. {\bf 291}, 167 (1985).


\bibitem{LiSch} L. Lindblom, M. Scheel,  {\em ``Energy Norms and 
the Stability of the Einstein Evolution Equations,''}
arXiv:gr-qc/0206035 

\bibitem{st} O. Sarbach and M. Tiglio, {\em ``Exploiting gauge and
constraint freedom in hyperbolic formulations of Einstein's
equations, } to appear in Phys. Rev. {\bf D}, arXiv:gr-qc/0205086.


\bibitem{c_match} J. Winicour, {\em Characteristic evolution and
matching}, Living Reviews in Relativity, arXiv:gr-qc/0102085. 

\bibitem{p_match} L. Rezzolla {\it et al.}, Phys. Rev. D {\bf 59}, 064001
(1999);  M. E. Rupright {\it et al.}, Phys. Rev. D {\bf 58}, 044005 (1998);
A. Abrahams {\it et al.}, Phys. Rev. Lett. {\bf 80}, 1812 (1998).

\end{thebibliography}
\end{document}